\documentclass[aps,reprint,nofootinbib]{revtex4-1}
\usepackage{graphicx,amsfonts,amsmath,amssymb}
\usepackage{bm}
\usepackage{url}
\makeatletter
\g@addto@macro{\UrlBreaks}{\UrlOrds}
\makeatother

\begin{document}
\title{A Jacobi Algorithm in Phase Space: Diagonalizing (skew-) 
Hamiltonian and Symplectic Matrices with Dirac-Majorana Matrices}
\date{\today}
\author{C. Baumgarten}
\affiliation{Paul Scherrer Institute, Switzerland}
\email{christian.baumgarten@psi.ch}

\def\begeq{\begin{equation}}
\def\endeq{\end{equation}}
\def\begary{\begeq\begin{array}}
\def\endary{\end{array}\endeq}
\newcommand{\myarray}[1]{\begin{equation}\begin{split}#1\end{split}\end{equation}}
\def\bmtx{\left(\begin{array}}
\def\emtx{\end{array}\right)}
\def\d{\partial}
\def\h{\eta}
\def\w{\omega}
\def\W{\Omega}
\def\s{\sigma}
\def\eps{\varepsilon}
\def\e{\epsilon}
\def\a{\alpha}
\def\b{\beta}
\def\g{\gamma}
\def\G{\Gamma}
\def\y{\gamma}
\def\d{\partial}
\def\S{{\Sigma}}

\def\leftD#1{\overset{\leftarrow}{#1}}
\def\rightD#1{\overset{\rightarrow}{#1}}

\begin{abstract}
Jacobi's method is a well-known algorithm in linear algebra to
diagonalize symmetric matrices by successive elementary rotations. 
We report here about the generalization of these elementary rotations
towards canonical transformations acting in Hamiltonian phase spaces. 
This generalization allows to use Jacobi's method in order to compute 
eigenvalues and eigenvectors of Hamiltonian (and skew-Hamiltonian) 
matrices with either purely real or purely imaginary eigenvalues by 
successive elementary ``decoupling'' transformations.
\end{abstract}

% 11.30.Cp Lorentz and Poincaré invariance
% 03.30.+p Special relativity
% 02.20.Hj Classical groups
\pacs{11.30.Cp,03.30.+p,02.20.Hj}
\keywords{Lorentz invariance, Special relativity, Clifford Algebras, 
Classical groups}
\maketitle

\begin{quotation}
The real importance of Einstein's work was that he introduced
Lorentz transformations as something fundamental in physics -- P.A.M. Dirac~\cite{DiracLT}
\end{quotation}

%%%%%%%%%%%%%%%%%%%%%%%%%%%%%%%%%%%%%%%%%%%%%%%%%%%%%%%%%%%%%%%%%%%%%%%%%%%%%%%%
\section{Introduction}
%%%%%%%%%%%%%%%%%%%%%%%%%%%%%%%%%%%%%%%%%%%%%%%%%%%%%%%%%%%%%%%%%%%%%%%%%%%%%%%%

The problem of eigenvector and eigenvalue computation of (skew-) symmetric,
Hamiltonian and symplectic matrices received considerable attention in the past
\footnote{See Refs.~\cite{LaubMeyer,GV,FMM,KS,MSW,BFS,MJB} and references therein.}. 
Here we describe a method that is entirely base on pure Hamiltonian (symplectic)
notions. We shall develop the real Clifford algebra $Cl(3,1)$ from algebraic
Hamiltonian symmetries and prove a morphism between the group of linear symplectic 
transformations in classical phase space and the Lorentz group.

Jacobi's Method is a well known numerical method that allows to diagonalize
symmetric matrices\footnote{See for instance Sec. 9.2 in Ref.~\cite{Scherer}.}.
The method is most easily explained when we start with some arbitrary real
symmetric $2\times 2$-matrix ${\bf A}$:
\begeq
{\bf A}=\bmtx{cc}a_{11}&a_{12}\\a_{12}&a_{22}\emtx
\endeq
This matrix can be diagonalized by an orthogonal Matrix ${\bf R}^T={\bf R}^{-1}$ of the form
\begeq
{\bf R}=\bmtx{cc}c& -s\\ s& c\emtx
\endeq
where $c=\cos{\theta}$ und $s=\sin{\theta}$, so that 
\myarray{
{\bf\tilde A}&={\bf R}\,{\bf A}\,{\bf R}^{-1}\\
\tilde a_{11}&=a_{11}\,c^2+a_{22}\,s^2-2\,a_{12}\,s\,c\\
\tilde a_{12}&=a_{12}\,(c^2-s^2)+(a_{11}-a_{22})\,s\,c\\
\tilde a_{22}&=a_{22}\,c^2+a_{11}\,s^2+2\,a_{12}\,s\,c\\
}
where $\tilde a_{12}$ simplifies to:
\begeq
\tilde a_{12}=a_{12}\,C+{a_{11}-a_{22}\over 2}\,S
\endeq
where $S=2\,s\,c=\sin{2\,\theta}$ and $C=c^2-s^2=\cos{2\,\theta}$ .
The transformed matrix ${\bf\tilde A}$ is diagonal, if $\tilde a_{12}=0$
and hence, if the angle of rotation $\theta$ is:
\myarray{
0&=a_{12}\,C+{a_{11}-a_{22}\over 2}\,S\\
\theta&=\frac{1}{2}\,\arctan{({2\,a_{12}\over a_{22}-a_{11}})}\\
}
This method allows to diagonalize real symmetric $n\times n$-matrices
by successive rotations, if the orthogonal rotation matrix ${\bf R}_{kl}$ is
a unit matrix except for the $k$-th and $l$-th rows and column. This
means that $r_{ii}=1$ for $i\notin {l,k}$ and $r_{ll}=r_{kk}=c$. The only
non-vanishing off-diagonal elements are $-r_{lk}=r_{kl}=s$:
\begin{equation}
{\bf R}=\begin{bmatrix}
\,\,{\bf 1} &        &    &       &    &  & \\
        & \ddots &    &       &    &  & \\
        &        &  c &       & -s &  & \\
        &        &    &\ddots &    &  & \\
        &        &  s &       &  c &  & \\
        &        &    &       &    &  \ddots & \\
        &        &    &       &    &         & {\bf 1}\,\,\\
\end{bmatrix}
\label{eq_Rmatrix}
\end{equation}
This rotation then allows to zero any selected non-diagonal element $a_{kl}=0$. 
The Jacobi method requires to chose the order of the successive rotations such 
that always the dominant non-diagonal element $a_{kl}$ is selected for the next 
rotation -- until convergence. The successive rotations can be collected in a 
matrix ${\bf R}$
\begeq
{\bf R}={\bf R}_{N},{\bf R}_{N-1},\dots,{\bf R}_{1}
\endeq
so that
\myarray{
{\bf\tilde A}&=\rm{Diag}(\alpha_1,\dots,\alpha_n)\\
             &={\bf R}\,{\bf A}\,{\bf R}^T\\
{\bf A}&={\bf R}^T\,{\bf\tilde A}\,{\bf R}\,.
}
Since real symmetric matrices have real eigenvalues, all $\alpha_i$ are
real.

\section{Phase Space and Hamiltonian Matrices}

The conventional Jacobi method operates on matrices in $n$-dimensional 
Euklidean space in the sense that orthogonal transformations do not
change the $n$-dimensional Euklidean norm. In contrast to geometrical
spaces ($\mathbb{R}^n$), phase spaces are dynamical spaces and the  
genuine transformations in phase spaces are canonical transformations,
as they preserve Hamilton's equations of motion.

Given a coordinate $\psi=(q_1,p_1,\dots,q_n,p_n)^T$ in a classical 
phase space of $n$ degrees of freedom, we define a Hamiltonian 
function ${\cal H}(\psi)$ by the quadratic form~\footnote{It suffices
to refer to the {\it quadratic} form of a more general Taylor series
of ${\cal H}(\psi)$ since we are only interested in linear transformations.}
\begeq
{\cal H}(\psi)=\frac{1}{2}\,\psi^T\,{\bf A}\,\psi 
\endeq
with a $2\,n\times 2\,n$-dimensional real symmetric matrix ${\bf A}$.
Hamilton's equations of motion are obtained from the condition
\begeq
\dot{\cal H}(\psi)=\sum\limits_{k=1}^{n}\,{\d{\cal H}\over\d q_k}\,\dot q_k+{\d{\cal H}\over\d p_k}\,\dot p_k=0
\endeq
with the general (but not unique) solution
\myarray{
\dot q_k&={\d{\cal H}\over\d p_k}\\
\dot p_k&=-{\d{\cal H}\over\d q_k}\\
}
This can be written in vectorial form as
\myarray{
\dot{\cal H}(\psi)&=\nabla_\psi\,{\cal H}(\psi)\cdot\dot\psi=0\\
\dot\psi&=\y_0\,\nabla_\psi\,{\cal H}(\psi)\\
        &=\y_0\,{\bf A}\,\psi\\
}
where the constraint that ${\cal H}(\psi)=\rm{const}$ leads to:
\begeq
\dot{\cal H}(\psi)=\psi^T\,{\bf A}\,\y_0\,{\bf A}\,\psi=0
\endeq
which is true whenever ${\bf A}\,\y_0\,{\bf A}$ is skew-symmetric,
i.e. whenever $\y_0$ is skew-symmetric. Typically the so-called
{\it symplectic unit matrix} (SUM) $\y_0$ is chosen to either have the form
\begeq
\y_0=\rm{Diag}(\eta,\eta,\dots,\eta)={\bf 1}_{n\times n}\otimes\eta
\endeq
with the $2\times 2$ block-diagonals
\begeq
\eta=\bmtx{cc}
0&1\\
-1&0\emtx
\endeq
or
\begeq
\y_0=\eta\,\otimes\,{\bf 1}_{n\times n}=\bmtx{cccc}0&{\bf 1}_{n\times n}\\
-{\bf 1}_{n\times n}&0\emtx\,.
\endeq
The former case corresponds to an ordering of variables in canonical pairs,
($\psi=(q_1,p_1,q_2,p_2,\dots)^T$), the latter case to a geometric ordering, 
i.e. first the coordinates and then the momenta: $\psi=(q_1,q_2,\dots,q_n,p_1,p_2,\dots,p_n)^T$.
The difference is mostly notational, but the algorithm that we aim to 
describe, favours the more dynamically motivated former case.

The evolution of the system is, for ${\bf A}=\rm{const}$ given by
the matrix exponential:
\begeq
\psi(t)=\exp{({\bf H}\,\tau)}\,\psi(0)\,.
\label{eq_solution}
\endeq
The matrix ${\bf H}$ that appears in the exponent 
\begeq
{\bf H}=\y_0{\bf A}
\endeq
is called a Hamiltonian matrix and it obeys the relation
\begeq
{\bf H}^T=\y_0\,{\bf H}\,\y_0
\label{eq_symplex}
\endeq
The simplest case is certainly that of a diagonal matrix ${\bf A}$,
and the unit matrix is the simplest diagonal matrix. Hence, if we
consider this simplest case, then ${\bf A}=\w\,{\bf 1}$, where $\w$
is the (eigen-) frequency. Then Eq.~\ref{eq_solution} results in
\begeq
\psi(t)=\exp{(\y_0\,\w\,\tau)}\,\psi(0)=\sum\limits_k\,{(\y_0\,\w\,\tau)^k\over k!}\,\psi(0)\,.
\endeq
which can be splitted in the even and odd powers:
\myarray{
\exp{(\y_0\,\w\,\tau)}&=\sum\limits_k\,{(\y_0\,\w\,\tau)^{2k}\over 2k!}+\sum\limits_k\,{(\y_0\,\w\,\tau)^{2k+1}\over (2k+1)!}\\
&=\sum\limits_k\,{(-)^k\,(\w\,\tau)^{2k}\over 2k!}+\y_0\,\sum\limits_k\,{(-)^k(\w\,\tau)^{2k+1}\over (2k+1)!}\\
&=\cos{(\w\,\tau)}+\y_0\,\sin{(\w\tau)}\,.
}
In this derivation we made use of nothing but the fact that $\y_0^2=-{\bf 1}$.
Hence, the matrix exponential of symmetric (Hamiltonian) matrices $\y_a$ with $\y_a^2=1$ yields 
correspondingly:
\myarray{
\exp{(\y_a\,\w\,\tau)}&=\sum\limits_k\,{(\y_a\,\w\,\tau)^{2k}\over 2k!}+\sum\limits_k\,{(\y_a\,\w\,\tau)^{2k+1}\over (2k+1)!}\\
&=\sum\limits_k\,{(\w\,\tau)^{2k}\over 2k!}+\y_a\,\sum\limits_k\,{(\w\,\tau)^{2k+1}\over (2k+1)!}\\
&=\cosh{(\w\,\tau)}+\y_a\,\sinh{(\w\tau)}\,.
\label{eq_mtxexp}
}
It is well-known (and easily proven) in linear Hamiltonian theory that the matrix exponentials 
of Hamiltonian matrices are symplectic and the matrix logarithm of any symplectic matrix is 
Hamiltonian~\cite{MHO}. A symplectic matrix ${\bf M}$ represents a linear canonical transformation 
of the phase space coordinates and fulfills the following definition:
\begeq
{\bf M}\,\y_0\,{\bf M}^T=\y_0\,,
\label{eq_symplectic}
\endeq
and the corresponding transformation is
\myarray{
\tilde\psi&={\bf M}\,\psi\\
\dot{\tilde\psi}&={\bf M}\,{\bf H}\,{\bf M}^{-1}\,\tilde\psi\\
}
Note that the matrix $\y_0$ is both, Hamiltonian {\it and} symplectic.

From Eq.~\ref{eq_symplectic} one obtains:
\begeq
{\bf M}^{-1}=-\y_0\,{\bf M}^T\,\y_0\,,
\label{eq_symplectic_inverse}
\endeq
so that the transformed Hamiltonian ${\bf\tilde H}$ is given by
\begeq
{\bf\tilde H}={\bf M}\,{\bf H}\,{\bf M}^{-1}\,.
\endeq
and must be Hamiltonian again, i.e.:
\myarray{
{\bf\tilde H}^T&=({\bf M}\,{\bf H}\,{\bf M}^{-1})^T\\
               &=({\bf M}^{-1})^T\,{\bf H}^T\,{\bf M}^T\\
               &=-(\y_0\,{\bf M}^T\,\y_0)^T\,{\bf H}^T\,{\bf M}^T\\
               &=-\y_0^T\,{\bf M}\,\y_0^T\,\y_0\,{\bf H}\,\y_0\,{\bf M}^T\\
               &=\y_0\,({\bf M}\,{\bf H}\,{\bf M}^{-1})\,\y_0\\
               &=\y_0\,{\bf\tilde H}\,\y_0\\
}
where Eq.~\ref{eq_symplectic} has been used. This means that symplectic
transformations are structure preserving: The form and structure of all
matrices and specifically the form of Hamilton's equations of motion
is preserved under symplectic similarity transformations.

\section{The real Pauli Algebra}
\label{sec_rpm}

The canonical pair is the smallest meaningful element in phase space
and accordingly the smallest Hamiltonian matrix has size $2\times 2$:
\begeq
{\bf H}=\bmtx{cc} h_{11}&h_{12}\\h_{21}&h_{22}\emtx
\endeq
The trace of the product of two matrices holds
\begeq
\rm{Tr}({\bf A}^T\,{\bf B})=\rm{Tr}({\bf A}\,{\bf B}^T)
\label{eq_trace}
\endeq
Hence, if ${\bf A}$ is symmetric and ${\bf B}$ skew-symmetric, then
the left side of Eq.~\ref{eq_trace} gives 
\begeq
\rm{Tr}({\bf A}^T\,{\bf B})=\rm{Tr}({\bf A}\,{\bf B})
\label{eq_trace1}
\endeq
and the right side
\begeq
\rm{Tr}({\bf A}\,{\bf B}^T)=-\rm{Tr}({\bf A}\,{\bf B})
\label{eq_trace2}
\endeq
which can only be true, if the trace of such a product vanishes:
The trace of the product of some symmetric ${\bf A}$ and some skew-symmetric 
matrix ${\bf B}$ is zero:
\begeq
\rm{Tr}({\bf A}\,{\bf B})=0
\label{eq_trace3}
\endeq
Hence all Hamiltonian matrices have a vanishing trace, i.e. $h_{11}+h_{22}=0$, 
so that a Hamiltonian $2\times 2$ matrix has the general form
\begeq
{\bf H}=\bmtx{cc} h_{11}&h_{12}\\h_{21}&-h_{11}\emtx
\endeq
Furthermore it follows from Eq.~\ref{eq_symplex} that symmetries play an essential
role in linear Hamiltonian theory. If a general Hamiltonian matrix is splitted
into it's symmetric ${\bf H}_s$ and skew-symmetric ${\bf H}_a$ parts
\myarray{
{\bf H}_s&=\frac{1}{2}\,({\bf H}+{\bf H}^T)\\
         &=\frac{\y_0}{2}\,({\bf H}\,\y_0-\y_0\,{\bf H})\\
{\bf H}_a&=\frac{1}{2}\,({\bf H}-{\bf H}^T)\\
         &=-\frac{\y_0}{2}\,({\bf H}\,\y_0+\y_0\,{\bf H})\,,
\label{eq_asym}
}
then it is evident that symmetries and (anti-) commutation
properties are related: the part of ${\bf H}$ which commutes 
with $\y_0$, is skew-symmetric and the part of ${\bf H}$ which
anti-commutes with $\y_0$, is symmetric. 
Hence it is required to distinguish symmetric from skew-symmetric components:
\begeq
{\bf H}=h_0\,\bmtx{cc} 0&1\\-1&0\emtx+h_1\,\bmtx{cc} 0&1\\1&0\emtx+h_2\,\bmtx{cc} 1&0\\0&-1\emtx
\label{eq_2x2}
\endeq
In consequence the parameters $h_1$, $h_2$ and $h_3$ represent dynamical structures with 
specific well-defined symmetry properties.

Since any $2\,n\times 2\,n$ Hamiltonian matrix may depend on $2\,n\,(2\,n+1)/2$ parameters,
$3$ parameters suffice to determine the form of $2\times 2$ Hamiltonian matrices
and Eq.~\ref{eq_2x2} is the most natural and convenient way to express this form: It is 
the form of the (matrix representation of a) Clifford algebra $Cl(1,1)$, namely the (real) 
Pauli algebra. The matrices appearing in Eq.~\ref{eq_2x2} are the real Pauli matrices:
\myarray{
\y_0=\eta_0&=\bmtx{cc} 0&1\\-1&0\emtx\\
\eta_1&=\bmtx{cc} 0&1\\ 1&0\emtx\\
\eta_2=\eta_0\,\eta_1&=\bmtx{cc} 1&0\\ 0&-1\emtx\\
}
The Pauli matrices mutually anti-commute and square to $\pm{\bf 1}$. In order to make this
system of matrices $\eta_i$ complete (i.e. a group), we define $\eta_3={\bf 1}_{2\times 2}$.

\section{The real Dirac Algebra}
\label{sec_rdm}

The original Jacobi algorithm picks out two ``spatial coordinates'' of 
$\mathbb{R}^n$ and ``rotates them'' by the use of $2\times 2$ orthogonal 
matrices. Since a canonical pair is the smallest meaningful element 
of phase space, a Jacobi algorithm for phase space must pick two 
canonical pairs and hence requires $4\times 4$ symplectic transformation 
matrices. The general form is then similar to
Eq.~\ref{eq_Rmatrix}, but the non-zero matrix elements $c$ and $s$ are
replaced by $2\times 2$-matrices. In analogy to Jacobi's algorithm
the objective of the algorithm is to achieve $2\times 2$ block-diagonalization, 
i.e. to reduce a $4\times 4$-problem to two ``decoupled'' (and trivial) 
$2\times 2$-problems:
\begeq
{\bf\tilde H}={\bf R}\,{\bf H}\,{\bf R}^{-1}=\bmtx{cc}
{\bf G}_{2\times 2}&0\\
0&{\bf K}_{2\times 2}\\
\emtx
\label{eq_bdiag}
\endeq
Let us first remark, before going into the details, that Hamiltonian
matrices have in general the following property: If $\lambda$ is an
eigenvalue of some Hamiltonian matrix ${\bf H}$, in the general case 
being complex, then $-\lambda$ as well as the complex conjugate 
values $\pm\bar\lambda$ are also eigenvalues of ${\bf H}$~\cite{MHO}. 
Since $2\times 2$ matrices have only two eigenvalues, $2\times 2$ Hamiltonian
matrices must have either purely real or purely imaginary eigenvalues.
Hence, if the Hamiltonian matrix ${\bf H}$ has complex eigenvalues off
axis (i.e. neither purely real nor purely imaginary), then 
symplectic block-diagonalization is impossible since similarity 
transformations, regardless whether they are symplectic or not, 
preserve the eigenvalues.

Furthermore, let us remark that $2\times 2$ Hamiltonian matrices
square to multiples of the unit matrix, since the real Pauli 
matrices mutually anti-commute:
\myarray{
{\bf H}^2&=(h_0\,\eta_0+h_1\,\eta_1+h_2\,\eta_2)^2\\
         &=-(h_0^2-h_1^2-h_2^2)\,{\bf 1}_{2\times 2}\\
\label{eq_H2x2sqr}
}
Hence the eigenvalues $\lambda$ of ${\bf H}_{2\times 2}$ are 
readily obtained by $\lambda=\pm\,\sqrt{-h_0^2+h_1^2+h_2^2}$.
Eq.~\ref{eq_H2x2sqr} provides evidence that the Hamiltonian matrix
of an oscillatory degree of freedom, normalized to the frequency, 
squares to $-1$ and is in this respect a generalization of the
unit imaginary. This likely is the reason why Hermann Weyl, who 
introduced the symplectic group, first used the term ``complex group''~\cite{Weyl}.

Since also real $4\times 4$ matrices can be written in terms of a 
Clifford algebra and since it was shown in Eq.~\ref{eq_mtxexp} 
that matrices which square to $\pm{\bf 1}$ not only have rather 
simple matrix exponentials\footnote{The computation of matrix exponentials is
in general significantly more involved~\cite{exp_paper}.} 
but also unique symmetry properties, it is nearby to 
use a Clifford algebraic parameterization of the required $4\times 4$ 
matrices. These can be obtained from the Kronecker products of the 
real Pauli matrices.

The Kronecker product of two matrices ${\bf A}=\{a_{ij}\}$ and ${\bf
  B}=\{b_{kl}\}$ is given by:
\begary{rcl}
{\bf C}&=&{\bf A}\otimes{\bf B}=\bmtx{cc}a_{11}{\bf B}&a_{12}{\bf B}\\a_{21}{\bf B}&a_{22}{\bf B}\emtx\\
&=&\bmtx{cccc}
a_{11}b_{11}&a_{11}b_{21}&a_{12}b_{11}&a_{12}b_{12}\\
a_{11}b_{12}&a_{11}b_{22}&a_{12}b_{21}&a_{12}b_{22}\\
a_{21}b_{11}&a_{21}b_{12}&a_{22}b_{11}&a_{22}b_{12}\\
a_{21}b_{21}&a_{21}b_{22}&a_{22}b_{21}&a_{22}b_{22}\\
\emtx\,,
\endary
i.e. the Kronecker product is a method to systematically write down all 
possible products between all elements of ${\bf A}$ and ${\bf B}$, respectively.
The rules of Kronecker matrix products are~\cite{MatrixAlgebra,VanLoan}:
\myarray{
({\bf A}\otimes{\bf B})^T&={\bf A}^T\otimes{\bf B}^T\\
{\bf A}\,\otimes\,({\bf B}+{\bf C})&={\bf A}\,\otimes\,{\bf B}+{\bf A}\,\otimes\,{\bf C}\\
({\bf A}\,\otimes\,{\bf B})\,({\bf C}\,\otimes\,{\bf D})&={\bf A}\,{\bf C}\otimes\,{\bf B}\,{\bf D}\\
\mathrm{Tr}({\bf A}\,\otimes\,{\bf B})&=\mathrm{Tr}({\bf A})\,\mathrm{Tr}({\bf B})\\
({\bf A}\otimes{\bf B})^{-1}&={\bf A}^{-1}\otimes{\bf B}^{-1}\\
\label{eq_Kronecker}
}
If some set $\eta_k$ of matrices represents a Clifford algebra, as in case of the 
real Pauli matrices, then it is straightforward to verify that the 
set of all Kronecker products $\eta_i\otimes\eta_j$ is again a representation
of some Clifford algebra~\footnote{More generally one finds (without proof) that 
all real matrix representations of Clifford algebras can be generated from (multiple)
Kronecker products of the real Pauli matrices.}.

Hence the required set of sixteen $4\times 4$ matrices is readily constructed 
from Kronecker products of real Pauli matrices. Let's label these matrices 
$\y_k$ with $k\in[0,\dots,15]$ where the (skew-symmetric) symplectic unit matrix 
(SUM) is $\y_0$ and the unit matrix is $\y_{15}={\bf 1}_{4\times 4}$. In the following
we shall investigate whether one can bring the remaining matrices into some
natural order.

As well known, the elements of Clifford algebras are generated 
(from products) of their mutually anti-commuting basis elements. 
A Clifford algebra of dimension $N$ has $N=p+q$ basis elements, 
called ``vectors'', $p$ of which square to ${\bf 1}$ (i.e. have 
signature $1$) and $q$ square to $-{\bf 1}$ (i.e. have signature $-1$). 
Products of two basis elements are ``bi-vectors'', products of $3$ 
elements are called ``tri-vectors'' and so on. The maximal vector 
degree is $N$ which is the product of all basis vectors and is 
called ``pseudo-scalar''. Real matrices of a given dimension 
$2^m\times 2^m$ always represent {\it some} Clifford algebra, 
but can only represent CAs of maximal possible dimension 
$N=p+q=2\,m$, if
\begeq
p-q=0,\,1,\,2\,\rm{ mod }\,8\,,
\endeq
which, since $N=p+q$ is even in our case, reduces to
\begeq
p-q=0,\,2\,\rm{ mod }\,8\,.
\label{eq_bott0}
\endeq
This is usually called Bott's periodicity~\cite{Bott,ABS,Okubo}.

The $N$ generators (or vectors) can be used to obtain new elements 
by multiplication since products of two (or more) {\it different} 
generators $\y_i\,\y_j$ are unique elements, different from the unit 
element and different from each factor. It follows from combinatorics 
that there are $\left({N\atop k}\right)$ $k$-vectors, so that one has
\begeq
\sum_k \left({N\atop k}\right)=2^N
\label{eq_pascal}
\endeq
elements in total. This means that the structure of Clifford algebras
is closely related to Pascal's triangle.

The question then is whether and how the Clifford algebraic structure might
support the objective to develop a Jacobi method for phase space. In order to
clarify this structure, let us first remark that all elements of any (real 
rep of a) Clifford algebra either commute or anti-commute with any other 
element, that any element is either purely symmetric or skew-symmetric and 
that {\it therefore} any element is either Hamiltonian or skew-Hamiltonian.

If we denote Hamiltonian elements by ${\bf S}_k$ and skew-Hamiltonian
elements by ${\bf C}_k$, then it is straightforward to show that
\begary{ccc}
\left.\begin{array}{c}
{\bf S}_1\,{\bf S}_2-{\bf S}_2\,{\bf S}_1\\
{\bf C}_1\,{\bf C}_2-{\bf C}_2\,{\bf C}_1\\
{\bf C}\,{\bf S}+{\bf S}\,{\bf C}\\
{\bf S}^{2\,n+1}\\
\end{array}\right\} & \Rightarrow & \mathrm{Hamiltonian}\\&&\\
\left.\begin{array}{c}
{\bf S}_1\,{\bf S}_2+{\bf S}_2\,{\bf S}_1\\
{\bf C}_1\,{\bf C}_2+{\bf C}_2\,{\bf C}_1\\
{\bf C}\,{\bf S}-{\bf S}\,{\bf C}\\
{\bf S}^{2\,n}\\
{\bf C}^n\\
\end{array}\right\} & \Rightarrow & \mathrm{skew-Hamiltonian}\\
\label{eq_cosy_algebra}
\endary
Hence the commutator of Hamiltonian elements is again Hamiltonian.

From Eq.~\ref{eq_asym} we know that any Clifford element that
commutes with $\y_0$, is skew-symmetric (with square $-{\bf 1}$)
and any component of ${\bf H}$ that anti-commutes with $\y_0$, is 
symmetric (with square to $+{\bf 1}$). 
By definition all elements of a Clifford basis mutually
anti-commute. Hence, if $\y_0$ is a basis element, then
all other basis elements {\it must} anti-commute with $\y_0$.
It follows that all other basis elements are either
Hamiltonian and symmetric or skew-Hamiltonian and skew-symmetric.

A purely Hamiltonian basis is hence only possible for Clifford
algebras of type $Cl(N-1,1)$ so that Bott's periodicity (Eq.~\ref{eq_bott0}
results in\footnote{
From a physical point of view, this might be an essential insight: 
It follows that Hamiltonian notions applied to real Clifford 
algebras {\it require} vector spaces in which the metric 
is of the Minkowski type.}:
\begeq
(N-1)-1=N-2=0,\,2\,\rm{ mod }\,8\,.
\label{eq_bott1}
\endeq
This includes the real Dirac algebra with $N=4$.
Hence a pure Hamiltonian basis exists and generates a unique 
representation of $Cl(3,1)$. Then there are $4$ Hamiltonian 
vector elements that form the Clifford basis and by 
Eq.~\ref{eq_pascal} we know that there are $\left({4\atop 2}\right)=6$ 
bi-vectors, which are, according to Eq.~\ref{eq_cosy_algebra}, 
all Hamiltonian.

It is straightforward to verify that products of two anti-commuting 
symmetric elements are skew-symmetric:
\begeq
(\y_1\,\y_2)^T=\y_2^T\,\y_1^T=\y_2\,\y_1=-\y_1\y_2\\
\endeq
while products of $\y_0$ and $\y_k$ ($k=1,2,3$) are symmetric 
bi-vectors. It follows that, when a SUM $\y_0$ and three more mutually 
anti-commuting symmetric basis elements $\y_1$, $\y_2$ and $\y_3$ are 
selected, the structure of the Hamiltonian Clifford algebra 
is completely determined. 

Since symmetric $4\times 4$-matrices have exactly $10=4+6$ 
linear independent parameters, the following structure emerges with
necessity~\footnote{To our knowledge, there is no such stringent logic 
to be found in complex representations of Clifford algebras.}:
1) There are $4$ basis elements, the SUM $\y_0$ and $3$ symmetric
elements $\y_1$, $\y_3$, $\y_3$.
2) There are $3$ symmetric bi-vectors: $\y_4=\y_0\,\y_1$, $\y_5=\y_0\,\y_2$,
and $\y_6=\y_0\,\y_3$.
3) There are $3$ skew-symmetric bi-vectors: $\y_7=\y_2\,\y_3$, $\y_8=\y_3\,\y_1$,
and $\y_9=\y_1\,\y_2$.
Note that $\y_0$ commutes with the skew-symmetric bi-vectors and
anti-commutes with the symmetric bi-vectors.

The remaining elements of the Clifford algebra are four $3$-vectors 
and the pseudo-skalar $4$-vector $\y_{14}=\y_0\y_1\y_2\y_3$, which 
are all skew-Hamiltonian. The skew-symmetric bi-vectors can also
be written as
\myarray{
\y_7&=\y_{14}\,\y_0\,\y_1\\
\y_8&=\y_{14}\,\y_0\,\y_2\\
\y_9&=\y_{14}\,\y_0\,\y_3\,.
}
It is sort of nearby to use a vectorial notation for the triple 
$\vec\y=(\y_1,\y_2,\y_3)^T$, so that we have the ``vectors'' 
$(\y_0,\vec\y)$ and the bi-vectors $\y_0\vec\y$ and 
$\y_{14}\y_0\vec\y$, respectively.
\begin{table}
\begin{tabular}{|l|c|c|c|}\hline
Type     & Symbol       & Symmetry & Hamiltonian\\\hline
1-vector & $\y_0$       & -    & +\\
         & $\vec\y$     & +    & +\\\hline
2-vector & $\y_0\vec\y$ & +    & +\\
         & $\y_{14}\y_0\vec\y$ & -    & +\\\hline
3-vector & $\y_{14}\,\y_0=\y_1\y_2\y_3$   &-  &-\\
         & $\y_{14}\,\vec\y$ &+  &-\\\hline
4-vector & $\y_{14}=\y_0\y_1\y_2\y_3$ &-&-\\
(pseudo-scalar)&&&\\\hline
skalar   & $\y_{15}={\bf 1}$ & + & -\\\hline
\end{tabular}
\caption[]{Structure of the symplectic Clifford algebra
$Cl(3,1)$. A plus indicates symmetry and a minus skewness.
\label{tab_rdms}
}
\end{table}
The question then is, whether and how the ten symplectic generators 
$\y_0,\dots,\y_9$ that have been identified, when regarded as generators
of symplectic transformations, allow to diagonalize real Hamiltonian 
$4\times 4$ matrices ${\bf H}$ and whether this provides the means for a 
symplectic Jacobi algorithm.

Let us therefore return to Eq.~\ref{eq_mtxexp} and investigate the effect
of transformations of the form 
\begeq
{\bf R}_i(\tau)=\exp{(\y_i\,\tau/2)}
\label{eq_transmtx}
\endeq~\footnote{
Note that the inverse transformation is given by the negative argument:
${\bf R}_i^{-1}={\bf R}(-\tau)=\exp{(-\y_i\,\tau/2)}$.} where $i\in[0,\dots,9]$:
\myarray{
{\bf R}_i\,\y_j\,{\bf R}_i^{-1}&=(c+\y_i\,s)\,\y_j\,(c-\y_i\,s)\\
&=c^2\,\y_j-\y_i\,\y_j\,\y_i\,s^2+(\y_i\,\y_j-\y_j\,\y_i)\,c\,s)\\
}
where $c=\cos{(\tau/2)}$ and $s=\sin{(\tau/2)}$ for $\y_i^2=-{\bf 1}$ and
$c=\cosh{(\tau/2)}$ and $s=\sinh{(\tau/2)}$ for $\y_i^2={\bf 1}$.
If $\y_i$ and $\y_j$ commute, this results in:
\begeq
{\bf R}_i\,\y_j\,{\bf R}_i^{-1}=\y_j\,,
\endeq
and if $\y_i$ and $\y_j$ anti-commute, one obtains:
\begeq
{\bf R}_i\,\y_j\,{\bf R}_i^{-1}=\y_j\,(c^2+\y_i^2\,s^2)+\y_i\,\y_j\,2\,c\,s\,.
\endeq
In case of a skew-symmetric $\y_i$, the final result is\footnote{
The half-angle arguments has been chosen in order to obtain full-angle
arguments here.
}:
\begeq
{\bf R}_i\,\y_j\,{\bf R}_i^{-1}=\y_j\,\cos{(\tau)}+\y_i\,\y_j\,\sin{(\tau)}\,.
\endeq
and in case of a symmetric $\y_i$:
\begeq
{\bf R}_i\,\y_j\,{\bf R}_i^{-1}=\y_j\,\cosh{(\tau)}+\y_i\,\y_j\,\sinh{(\tau)}\,.
\endeq
Hence skew-symmetric $\y_i$ generate rotation-like transformations and 
symmetric $\y_i$ generate boost-like symplectic transformations.

Note that we obtained the complete structure of the algebra from only 
two requirements, namely that $\y_0$ is a basis element of the Clifford
algebra and that {\it all} elements of the basis should be Hamiltonian
as defined by Eq.~\ref{eq_symplex}.
These two (nearby) conditions inevitably lead to the unique structure 
of $Cl(3,1)$ as listed in Tab.~\ref{tab_rdms}.

\section{The Electromagnetic Equivalence}
\label{sec_emeq}

Since any real-valued $4\times 4$-matrix ${\bf M}$ can be written as
\begeq
{\bf M}=\sum\limits_{k=0}^{15}\,m_k\,\y_k\,,
\endeq
any real-valued {\it Hamiltonian} $4\times 4$-matrix ${\bf H}$
can accordingly be written as:
\begeq
{\bf H}=\sum\limits_{k=0}^{9}\,h_k\,\y_k
\endeq
Since the unit element $\y_{15}={\bf 1}_{4\times 4}$ is the only $\y$-matrix
with non-vanishing trace, the trace of ${\bf M}$ is $4\,m_{15}$. 
The coefficients $m_k$ therefore can be obtained from the trace of the
product of $\y_k^T$ and ${\bf M}$
\begeq
m_k=\frac{1}{4}\,\rm{Tr}(\y_k^T\,{\bf M})\,,
\endeq
since $\y_k^T\,\y_k={\bf 1}$ for all $k\in[0,\dots,15]$.
Accordingly the coefficients of ${\bf H}$ are given by
\begeq
h_k=\frac{1}{4}\,\rm{Tr}(\y_k^T\,{\bf H})\,.
\endeq
The structure of $Cl(3,1)$ is, beyond the sign of the metric, closely
related to $Cl(1,3)$ which is used in the usual presentation of
the Dirac equation. Indeed the ``real Dirac matrices'' are, up to 
multiplication with the unit imaginary, identical to the Majorana
matrices~\cite{Majorana}. We shall not discuss here whether (and which)
physical insights this might eventually provide, but the transformation 
properties of the ten relevant parameters have a formal correspondence 
to (quantum-) electrodynamics and therefore it is convenient 
to use this morphism by the following notation~\cite{rdm_paper}:
\myarray{
  h_0&={\cal E}\\
  (h_1,h_2,h_3)^T&=\vec P\\
  (h_4,h_5,h_6)^T&=\vec E\\
  (h_7,h_8,h_9)^T&=\vec B\\
}
so that any $4\times 4$ Hamiltonian matrix can now be written as
the sum of the vector components ${\bf P}$ and the bi-vector components
${\bf F}$ as follows:
\myarray{
{\bf P}&={\cal E}\,\y_0+\vec P\cdot\vec\y={\cal E}\,\y_0+P_x\,\y_1+P_y\,\y_2+P_z\,\y_3\\
{\bf F}&=\y_0\,\vec E\cdot\vec\y+\y_{14}\y_0\vec B\cdot\vec\y\\
       &=E_x\,\y_4+E_y\,\y_5+E_z\,\y_6+B_x\,\y_7+B_y\,\y_8+B_z\,\y_9\\
{\bf H}&={\bf P}+{\bf F}\\
}
\begin{table}
\begin{tabular}{||c||c||c|c|c||}\hline
                  & $\y_0$   & $\y_7$ &  $\y_8$&  $\y_9$ \\\hline\hline
 $\tilde\y_0=\y_0$     &         &        &        &         \\\hline\hline
 $\tilde\y_1=c\,\y_1..$& $-s\,\y_4$&         & $-s\,\y_3$& $+s\,\y_2$ \\\hline
 $\tilde\y_2=c\,\y_2..$& $-s\,\y_5$& $+s\,\y_3$&        & $-s\,\y_1$ \\\hline
 $\tilde\y_3=c\,\y_3..$& $-s\,\y_6$& $-s\,\y_2$& $+s\,\y_1$&         \\\hline\hline
 $\tilde\y_4=c\,\y_4..$& $+s\,\y_1$&         & $-s\,\y_6$&  $\y_5$ \\\hline
 $\tilde\y_5=c\,\y_5..$& $+s\,\y_2$& $+s\,\y_6$&        & $-s\,\y_4$ \\\hline
 $\tilde\y_6=c\,\y_6..$& $+s\,\y_3$& $-s\,\y_5$& $+s\,\y_4$&         \\\hline\hline
 $\tilde\y_7=c\,\y_7..$&         &         & $-s\,\y_9$& $+s\,\y_8$ \\\hline
 $\tilde\y_8=c\,\y_8..$&         & $+s\,\y_9$&        & $-s\,\y_7$ \\\hline
 $\tilde\y_9=c\,\y_9..$&         & $-s\,\y_8$& $+s\,\y_7$&         \\\hline\hline
\end{tabular}
\caption{Table of symplectic rotations of two degrees of freedom. 
$a$ indicates the rows and $b$ the column:
$\y_a'=\exp{(\y_b\,{\tau/2})}\,\y_a\,\exp{(-\y_b\,{\tau/2})}$. If $\y_a$ and $\y_b$ anticommute, then the 
result is $\y_a'=c\,\y_a+s\,\y_a\,\y_b$ where $c$ and $s$ are the sine- and cosine-function of $\tau$.
\label{tab_symrot}}
\end{table}

\begin{table}
\begin{tabular}{||c||c|c|c||c|c|c||}\hline
                & $\y_1$  &  $\y_2$ &  $\y_3$&  $\y_4$   &  $\y_5$&  $\y_6$ \\\hline\hline
 $\tilde\y_0=c\,\y_0..$& $+s\,\y_4$& $+s\,\y_5$& $+s\,\y_6$& $-s\,\y_1$ & $-s\,\y_2$& $-s\,\y_3$\\\hline\hline
 $\tilde\y_1=c\,\y_1..$&         & $+s\,\y_9$& $-s\,\y_8$& $-s\,\y_0$ &        &        \\\hline
 $\tilde\y_2=c\,\y_2..$& $-s\,\y_9$&        & $+s\,\y_7$&         & $-s\,\y_0$&         \\\hline
 $\tilde\y_3=c\,\y_3..$& $+s\,\y_8$& $-s\,\y_7$&        &         &        & $-s\,\y_0$ \\\hline\hline
 $\tilde\y_4=c\,\y_4..$& $+s\,\y_0$&        &        &         & $+s\,\y_9$& $-s\,\y_8$ \\\hline
 $\tilde\y_5=c\,\y_5..$&        & $+s\,\y_0$&        &  $-s\,\y_9$&        & $+s\,\y_7$ \\\hline
 $\tilde\y_6=c\,\y_6..$&        &        & $+s\,\y_0$&  $+s\,\y_8$& $-s\,\y_7$&         \\\hline\hline
 $\tilde\y_7=c\,\y_7..$&        & $-s\,\y_3$& $+s\,\y_2$&         & $-s\,\y_6$& $+s\,\y_5$\\\hline
 $\tilde\y_8=c\,\y_8..$&$+s\,\y_3$&        & $-s\,\y_1$&  $+s\,\y_6$&        & $-s\,\y_4$\\\hline
 $\tilde\y_9=c\,\y_9..$&$-s\,\y_2$& $+s\,\y_1$&        &  $-s\,\y_5$& $+s\,\y_4$&         \\\hline\hline
\end{tabular}
\caption{Table of symplectic boosts with symmetric generator of two degrees of freedom. 
$a$ indicates the rows and $b$ the column: $\tilde\y_a=\exp{(\y_b\,{\tau/2})}\,\y_a\,\exp{(-\y_b\,{\tau/2})}$. 
If $\y_a$ and $\y_b$ anticommute, then the result is $\tilde\y_a=c\,\y_a+s\,\y_a\,\y_b$ where $c$ and $s$ 
are the hyperbolic sine- and cosine-function of $\tau$. If $\y_a$ and $\y_b$ commute, then $\y_a'=\y_a$.
\label{tab_symboost}}
\end{table}

With the basis given in App.~\ref{app_rdm} one obtains the following 
explicit form for the vectors:
{\small\begeq
{\bf P}=\bmtx{cccc}
-P_z&{\cal E}-P_x&0&P_y\\
-{\cal E}-P_x&P_z&P_y&0\\
0&P_y&-P_z&{\cal E}+P_x\\
P_y&0&-{\cal E}+P_x&P_z\\
\emtx
\label{eq_Pmtx}
\endeq}
For the bi-vectors one obtains:
{\small\begeq
{\bf F}=\bmtx{cccc}
-E_x&E_z+B_y&E_y-B_z&B_x\\
E_z-B_y&E_x&-B_x&-E_y-B_z\\
E_y+B_z&B_x&E_x&E_z-B_y\\
-B_x&-E_y+B_z&E_z+B_y&-E_x\\
\emtx
\label{eq_Fmtx}
\endeq}
The explicite computation of the effects of symplectic similarity 
transformations are given in Tab.~\ref{tab_symrot} and Tab.~\ref{tab_symboost}.
The calculations verify that the gyroscopic Hamiltonian terms
associated here with $\vec B$ generate rotations which are
isomorph to spatial rotation of vectors. Likewise the 
symmetric bi-vector terms $\vec E$ generate boosts that are
isomorphic to Lorentz-Boosts. This isomorphism is indeed helpful
to grasp the geometric content which is needed to obtain 
block-diagonalization.

\section{Block-Diagonalization}
\label{sec_blockdiag}

The analogy with the conventional Jacobi algorithm is achieved
by a block-diagonalization of ${\bf H}$. It is achieved when 
the transformed matrix 
\begeq
{\bf\tilde H}={\bf M}\,{\bf H}\,{\bf M}^{-1}
\endeq
has block-diagonal form, which means that $\tilde P_y=\tilde E_y=\tilde B_x=\tilde B_z=0$ 
(see Eq.~\ref{eq_Pmtx} and Eq.~\ref{eq_Fmtx}). In other words, the ``vectors''
$\vec P$ and $\vec E$ must be orthogonal to $\vec B$.
Let's define the following auxiliary ``scalar products''
\myarray{
\eps_r&=\vec E\cdot\vec B\\
\eps_g&=\vec B\cdot\vec P\\
\eps_b&=\vec E\cdot\vec P\\
\label{eq_aux_eps}
}
The first objective that has to be achieved, is therefore
$\eps_r=\eps_g=0$. The second objective is to rotate
the ``system'' such that $\vec B=B_y$.
We introduce the following auxiliary ``vectors'': 
\begary{rcl}
\vec r&\equiv&{\cal E}\,\vec P+\vec B\times \vec E \\
\vec g&\equiv&{\cal E}\,\vec E+\vec P\times \vec B \\
\vec b&\equiv&{\cal E}\,\vec B+\vec E\times \vec P \,,
\label{eq_aux_vecs}
\endary
The ``scalar products'' of Eq.~\ref{eq_aux_vecs} are invariant under spatial 
rotations. 
Hence we need to analyze the transformation behavior of these scalar products
only under symplectic boosts and under the (yet unnamed) rotation generated 
by $\y_0$. We introduce the following abbreviations for a better readability
\begary{rclp{5mm}rcl}
c&=&\cos{(\tau)} && s&=&\sin{(\tau)}\\
c_2&=&\cos{(2\,\tau)} && s_2&=&\sin{(2\,\tau)}\\
C&=&\cosh{(\tau)}&&S&=&\sinh{(\tau)}\\
C_2&=&\cosh{(2\,\tau)}&&S_2&=&\sinh{(2\,\tau)}\\
\label{eq_aux_angles}
\endary
\begin{table}
\begin{tabular}{|c||c|c|c|}\hline
             & $\tilde\eps_r$               & $\tilde\eps_g$             & $\tilde\eps_b$           \\\hline\hline
  $\y_0$     & $\eps_r\,c+\eps_g\,s$         & $\eps_g\,c-\eps_r\,s$     & $\eps_b\,c_2+{\vec P^2-\vec E^2\over 2}\,s_2$\\\hline\hline
  $\y_1$     & $\eps_r\,C-(\vec b)_x\,S$     & $\eps_g$                  & $\eps_b\,C-(\vec r)_x\,S$\\\hline
  $\y_2$     & $\eps_r\,C-(\vec b)_y\,S$     & $\eps_g$                  & $\eps_b\,C-(\vec r)_y\,S$\\\hline
  $\y_3$     & $\eps_r\,C-(\vec b)_z\,S$     & $\eps_g$                  & $\eps_b\,C-(\vec r)_z\,S$\\\hline\hline
  $\y_4$     & $\eps_r$                      & $\eps_g\,C+(\vec b)_x\,S$ & $\eps_b\,C+(\vec g)_x\,S$\\\hline
  $\y_5$     & $\eps_r$                      & $\eps_g\,C+(\vec b)_y\,S$ & $\eps_b\,C+(\vec g)_y\,S$\\\hline
  $\y_6$     & $\eps_r$                      & $\eps_g\,C+(\vec b)_z\,S$ & $\eps_b\,C+(\vec g)_z\,S$\\\hline
\end{tabular}
\caption{Table of ``scalar products'' after symplectic transformations generated by symplectic
boosts and $\y_0$, respectively (left column). 
See Eqns.~\ref{eq_aux_eps},~\ref{eq_aux_vecs} and~\ref{eq_aux_angles}. 
\label{tab_eps_trans}}
\end{table}
The results are summarized in Tab.~\ref{tab_eps_trans}. The inspection of
this tabel reveals that $\eps_r$ is invariant under the action of 
${\bf R}_{4,5,6}$ while $\eps_g$ is invariant under the action of ${\bf
R}_{1,2,3}$. This findings suggests the following possible strategies: 
Firstly, we can use ${\bf R}_0$ to either make $\eps_r=0$,
then (one of) ${\bf R}_{4,5,6}$ to make $\eps_g=0$ {\it or} we could first use ${\bf R}_0$ 
to make $\eps_g=0$, followed by ${\bf R}_{1,2,3}$ to make $\eps_r=0$.
In both cases, it is possible to use spatial rotations (${\bf R}_{7,8,9}$) to 
align $\vec b$ along one axis, preferably the $y$-axis since in the chosen 
matrix representation, $b_x=b_z=0$ will eventually lead to the required
``vector orientation'' of $\vec B=B_y$, i.e. provide $B_x=B_z=0$.

Hence we can identify four steps that are needed to achieve block-diagonalization.
After each step, the coefficients $h_k$, the auxiliary parameters $\eps_i$
and the ``vectors'' $\vec r,\vec g,\vec b$ have to be re-evaluated. 
The required steps are:
1) Rotate using ${\bf R}_0$ (see Eq.~\ref{eq_transmtx}) and angle 
$\tau=-\arctan{(\eps_r/\eps_g)}$. This result in $\tilde\eps_r=0$.
2) Rotate $\vec b$ using ${\bf R}_9$ with angle $\tau=-\arctan{(b_x/b_y)}$ such that $\tilde b_x=0$. 
3) Rotate $\vec b$ using ${\bf R}_7$ with angle $\tau=\arctan{(b_z/b_y)}$ such that $\tilde b_z=0$. 
4) The next step is to apply a boost with ${\bf R}_5$ and $\tau=-\rm{artanh}{(\eps_g/b_y)}$.
This last step requires that $\vert\eps_g/b_y\vert< 1$ (at this step). Otherwise the
Hamiltonian matrix cannot be block-diagonalized and has complex eigenvalues.
The block-diagonal matrix ${\bf\tilde H}$ is then given by:
\begeq
{\bf\tilde H}={\bf R}_5\,{\bf R}_7\,{\bf R}_9\,{\bf R}_0\,{\bf H}\,{\bf R}_0^{-1}\,{\bf R}_9^{-1}\,{\bf R}_7^{-1}\,{\bf R}_5^{-1}
\endeq

Alternatively we could also proceed as follows:
1) Rotate using ${\bf R}_0$ (see Eq.~\ref{eq_transmtx}) and angle 
$\tau=\arctan{(\eps_g/\eps_r)}$. This should make $\tilde\eps_g=0$.
2) Rotate $\vec b$ using ${\bf R}_9$ with angle $\tau=-\arctan{(b_x/b_y)}$ such that $\tilde b_x=0$. 
3) Rotate $\vec b$ using ${\bf R}_7$ with angle $\tau=\arctan{(b_z/b_y)}$ such that $\tilde b_z=0$. 
4) The last step is to apply a boost with ${\bf R}_2$ and $\tau=\rm{artanh}{(\eps_r/b_y)}$.
Again, this is only possible for $\vert\eps_r/b_y\vert< 1$ (at this step). 
In summary:
\begeq
{\bf\tilde H}={\bf R}_2\,{\bf R}_7\,{\bf R}_9\,{\bf R}_0\,{\bf H}\,{\bf R}_0^{-1}\,{\bf R}_9^{-1}\,{\bf R}_7^{-1}\,{\bf R}_2^{-1}
\endeq

In order to block-diagonalize $2\,n\times 2\,n$ Hamiltonian matrices, the algorithm follows
the same logic as the usual Jacobi algorithm and requires, before each step, to select the 
``dominant'' off-diagonal $2\times 2$-block to be ``treated'' next. We tested
the algorithm by selecting the $2\times 2$-block which has the largest sum of
squared elements. We found a convergence that depends quadratically on the
number of the degrees of freedom (i.e. $2\times 2$-blocks) $n$, i.e. with 
${\cal O}(n^2)$ steps~\cite{geo_paper}. We do not provide a rigorous proof of convergence. 

\section{Diagonalization}

After successfull block-diagonalization, the problem is reduced to
that of Hamiltonian $2\times 2$-matrices in form of Eq.~\ref{eq_bdiag} and
Eq.~\ref{eq_2x2}, respectively.
Eq.~\ref{eq_2x2} can further be simplified (if desired) by the use of the
following transformation:
\myarray{
{\bf H}_{2\times 2}&=h_0\,\eta_0+h_1\,\eta_1+h_2\,\eta_2\\
{\bf\tilde H}_{2\times 2}&=\exp{(\eta_0\,\tau/2)}\,{\bf H}_{2\times 2}\,\exp{(-\eta_0\,\tau/2)}\\
}
with $\tau=-\arctan{(h_2/h_1)}$ one obtains:
\myarray{
{\bf\tilde H}_{2\times 2}&=\bmtx{cc}
0&h_0+\tilde h_1\\
-h_0+\tilde h_1&0\\
\emtx\\
\tilde h_1&=h_1\,\sqrt{1+(h_2/h_1)^2}\\
}
In the next step, we transform using the generator $\eta_2$:
\myarray{
{\bf\tilde H}_{2\times 2}&\to\exp{(\eta_2\,\tau/2)}\,{\bf H}_{2\times 2}\,\exp{(-\eta_2\,\tau/2)}\\
&=\bmtx{cc}
0&\tilde h_0\\
-\tilde h_0&0\\
\emtx\\
\tau&=\log{\left(\sqrt{h_0-\tilde h_1}\over\sqrt{h_0+\tilde h_1}\right)}\\
\tilde h_0&=\w=\sqrt{h_0^2-h_1^2-h_2^2}\\
\label{eq_normalform}
}
This is still not a diagonal matrix, but it contains all required information
and it is all that can be done by the use of symplectic transformations.
The form of ${\bf\tilde H}_{2\times 2}$, as given in the first line of 
Eq.~\ref{eq_normalform}, is the normal form of a one-dimensional oscillator.
 
Since the eigenvalues of ${\bf\tilde H}_{2\times 2}$ are directly given 
as $\pm\,i\,\w$, it is possible -- though not required -- to take the next 
step and diagonalize the matrix. We provide this step only for 
completeness. The (almost trivial) matrix ${\bf V}$ of eigenvectors and 
the final transformation to diagonal form are:
\myarray{
{\bf V}&={1\over\sqrt{2}}\bmtx{cc}i&1\\-i&1\emtx\\
{\bf V}^{-1}&={1\over\sqrt{2}}\bmtx{cc}-i&i\\1&1\emtx\\
{\bf V}\,{\bf\tilde H}_{2\times 2}\,{\bf V}^{-1}&=\bmtx{cc}i\,\w&0\\0&-i\,\w\emtx\\
}
This last step is not a symplectic transformation. Nonetheless it is an
interesting step as it reveals the last (or first) step that transforms 
between a ``classical'' real and a seemingly ``non-classical'' 
(complex) form of Hamiltonian dynamics, which consists in replacement 
\myarray{
\tilde\psi&={\bf V}\,\psi={1\over\sqrt{2}}\bmtx{cc}i&1\\-i&1\emtx\,\bmtx{c}q\\p\emtx\\
          &={1\over\sqrt{2}}\,\bmtx{c}p+i\,q\\p-i\,q\emtx\\
}
The equations of motion then read, after multiplication with the unit
imaginary and $\hbar$:
\begeq
i\,\hbar\,\dot {\tilde\psi}=\bmtx{cc}-\hbar\w&0\\0&\hbar\w\emtx\,\tilde\psi\,,
\endeq

\section{Eigenvalues}

The eigenvalues of ${\bf H}_{4\times 4}$ can also be obtained by a different
method. It is well known in linear algebra that the trace of a matrix is
the sum of it's eigenvalues, the trace of the square of a matrix is the 
sum of squares of it's eigenvalues. Since the sum of the eigenvalues of
a Hamiltonian matrix (and it's odd powers) vanishes, the trace of
the even powers do not vanish. Recall that those Hamiltonian matrices 
that allow for block-diagonalization, have either purely real or purely
imaginary eigenvalues. Let $\pm\lambda$ and $\pm\Lambda$ be the four 
eigenvalues, then:
\myarray{
\rm{Tr}({\bf H}^2)&=2\,\lambda^2+2\,\Lambda^2\\
\rm{Tr}({\bf H}^4)&=2\,\lambda^4+2\,\Lambda^4\\
}
If we introduce the definitions 
\myarray{
K_1&=\rm{Tr}({\bf H}^2)/4={\lambda^2+\Lambda^2\over 2}\\
K_2&=\rm{Tr}({\bf H}^4)/16-K_1^2/4\\
   &={\lambda^4+\Lambda^4\over 8}-{\lambda^4+\Lambda^4+2\,\lambda^2\,\Lambda^2\over 16}\\
   &={(\lambda^2-\Lambda^2)^2\over 16}\\
}
so that one obtains:
\myarray{
K_1+2\,\sqrt{K_2}&={\lambda^2+\Lambda^2\over 2}+{\lambda^2-\Lambda^2\over 2}=\lambda^2\\
K_1-2\,\sqrt{K_2}&={\lambda^2+\Lambda^2\over 2}-{\lambda^2-\Lambda^2\over 2}=\Lambda^2\\
}
and hence the eigenvalues can be determined from $K_1$ and $K_2$ alone:
\myarray{
\lambda&=\pm\,\sqrt{K_1+2\,\sqrt{K_2}}\\
\Lambda&=\pm\,\sqrt{K_1-2\,\sqrt{K_2}}\\
}
Inserting Eq.~\ref{eq_Pmtx} and Eq.~\ref{eq_Fmtx} yields~\cite{osc_paper}:
\myarray{
K_1&=-{\cal E}^2-\vec B^2+\vec P^2+\vec E^2\\
K_2&=({\cal E}\,\vec B+\vec E\times\vec P)^2-(\vec E\cdot\vec B)^2-(\vec P\cdot\vec B)^2\\
   &=\vec b^2-\eps_r^2-\eps_g^2\\
}
There are two cases of special interest, namely if ${\bf H}={\bf P}$ (only vector components)
\myarray{
K_1&=-{\cal E}^2+\vec P^2\\
K_2&=0\,,
}
and ${\bf H}={\bf F}$ (only bi-vector components),
\myarray{
K_1&=\vec E^2-\vec B^2\\
K_2&=-(\vec E\cdot\vec B)^2\,.
}
Recall that symplectic eigenvalues are {\it invariants}.

\section{Symplectic Matrices}

According to linear Hamiltonian theory, every symplectic matrix ${\bf M}$
is the matrix exponentials of some Hamiltonian matrix ${\bf H}$~\cite{MHO}:
\myarray{
{\bf M}&=\exp{({\bf H}\,\tau)}\\
       &={\bf 1}+{\bf H}\,\tau+\frac{1}{2}\,{\bf H}^2\,\tau^2+\dots\,.
\label{eq_TaylorExp}
}
Since any analytic function of block-diagonal matrices is again a block-diagonal
matrix it is evident that ${\bf M}$ is blockdiagonal whenever ${\bf H}$ is block-
diagonal. From Eq.~\ref{eq_cosy_algebra} we know that all odd terms of the
Taylor series Eq.~\ref{eq_TaylorExp} are Hamiltonian and all even terms are
skew-Hamiltonian. Hence, we can use the brute force method and remove all 
skew-Hamiltonian terms from ${\bf M}$:
\begeq
{\bf\tilde M}=\frac{1}{2}\,({\bf M}+\y_0\,{\bf M}^T\,\y_0)\,.
\endeq
Whatever similarity transformation is used to block-diagonalize ${\bf\tilde M}$,
it will automatically block-diagonalize ${\bf M}$ as well.

In other words: If a Hamiltonian matrix can be written as 
\begeq
{\bf H}={\bf V}\,{\bf D}\,{\bf V}^{-1}\,,
\endeq
then the matrix exponential of ${\bf H}$ is:
\begeq
{\bf M}=\exp{({\bf H})}={\bf V}\,\exp{({\bf D})}\,{\bf V}^{-1}\,,
\endeq
so that an eigenvalue $\lambda$ of ${\bf H}$ is replaced by matrix exponentiation with
$e^\lambda$ in ${\bf M}$.

\section{Skew-Hamiltonian Matrices}

Skew-Hamiltonian matrices ${\bf C}$, when expressed with the real Dirac algebra, can
be written as (see Tab.~\ref{tab_rdms}):
\begeq
{\bf C}=\sum\limits_{k=10}^{15}\,c_k\,\y_k\,.
\endeq
The four components $(c_{10},c_{11},c_{12},c_{13})$ are sometimes called pseudo-vector or ``axial''
vector and they transform accordingly. The pseudo-scalar $c_{14}$ and the scalar $c_{15}$ are
invariants. For instance ${\bf H}^2/2$ is skew-Hamiltonian and has the following components:
\myarray{
c_{10}&=\vec P\,\cdot\,\vec B\\
(c_{11},c_{12},c_{13})^T&={\cal E}\,\vec B+\vec E\times\vec P\\
c_{14}&=\vec E\,\cdot\,\vec B\\
c_{15}&=(-{\cal E}^2+\vec P^2-\vec B^2+\vec E^2)/2\,.
}
The reader will have noticed that $c_{10}=\eps_g$, $c_{14}=\eps_r$ and $(c_{11},c_{12},c_{13})^T=\vec b$
have been defined before in Eq.~\ref{eq_aux_eps} and Eq.~\ref{eq_aux_vecs}. Hence the simplest
way to determine those parameters, is to compute ${\bf H}^2$ and extract those components:
\myarray{
\eps_g&=\rm{Tr}(\y_{10}^T\,{\bf H}^2)/8\\
\eps_r&=\rm{Tr}(\y_{14}^T\,{\bf H}^2)/8\\
b_x&=\rm{Tr}(\y_{11}^T\,{\bf H}^2)/8\\
b_y&=\rm{Tr}(\y_{12}^T\,{\bf H}^2)/8\\
b_z&=\rm{Tr}(\y_{13}^T\,{\bf H}^2)/8\\
}
Block-diagonalization requires that $c_{10}=c_{11}=c_{13}=c_{14}=0$, or, in the notation
introduced above that $\eps_r=\eps_g=b_x=b_z=0$. This is exactly the strategy of the 
described algorithm. Hence the developed algorithm is {\it already}
able to block-diagonalize skew-Hamiltonian matrices.

\section{Applications}

Many problems in engineering and physics involve Hamiltonian matrices, often of size 
greater than $2\times 2$. An example from accelerator physics
can, for instance, be found in Ref.~\cite{cyc_paper}. Since most accelerators
make use of a median plane symmetry, the vertical degree of freedom does, in
linear approximation, not couple to the transverse horizontal and the
longitudinal degrees of freedom of particles in a frame co-moving with the
bunch. Then the involved $6\times 6$ transfer matrices decay into two
problems, a $4\times 4$ and a $2\times 2$ problem so that, in this specific
case, no iteration is needed.

To model and understand the collective behavior of charged particles inside a 
bunch is of severe importance in accelerator physics.
The coordinates $\psi$ of individual particles are the dynamical variables
and they represent (small) {\it deviations} of a particle's position and momentum 
relative to some reference orbit. But the ``observables'' that are relevant to describe
the collective behavior are not individual particle positions but statistical averages
of particle ensembles as for instance the root-mean-square width of the beam,
which is (in square) an element of the matrix of second moments:
\begeq
\Sigma=\langle\psi\psi^T\rangle\,.
\endeq
The evolution of $\Sigma$ in time is given by
\myarray{
\dot\Sigma&=\langle\dot\psi\psi^T\rangle+\langle\psi\dot\psi^T\rangle\\
          &=\langle{\bf H}\psi\psi^T\rangle+\langle\psi\psi^T{\bf H}^T\rangle\\
          &={\bf H}\,\Sigma+\Sigma\,{\bf H}^T\\
          &={\bf H}\,\Sigma+\Sigma\,\y_0\,{\bf H}\,\y_0\,,
}
where ${\bf H}$ is the (local) Hamiltonian matrix.
If we multiply the last line with $\y_0^T$ from the right, we obtain:
\myarray{
\dot\Sigma\,\y_0^T&={\bf H}\,(\Sigma\y_0^T)-(\Sigma\,\y_0^T)\,{\bf H}\\
\dot{\bf S}&={\bf H}\,{\bf S}-{\bf S}\,{\bf H}\,,
\label{eq_Heisenberg}
}
with the Hamiltonian matrix ${\bf S}=\Sigma\y_0^T$. The last line tells us that 
${\bf H}$ and ${\bf S}$ are a so-called ``Lax pair''. As Peter Lax has proven\cite{Lax}, 
it follows from the validity of Eq.~\ref{eq_Heisenberg} that the following expression
\begeq
\rm{Tr}({\bf S}^k)=\rm{const}
\endeq
is invariant for any integer $k$.

Since the evolution of $\psi$ in time is given by the symplectic transformation 
\begeq
\psi(t)={\bf M}(t)\,\psi(0)=\exp{({\bf H}\,\tau)}\,\psi(0)\,,
\endeq
then we obtain for the the matrix of second moments:
\myarray{
\Sigma(t)&={\bf M}\,\Sigma(0)\,{\bf M}^T\\
{\bf S}(t)&={\bf M}\,{\bf S}(0)\,\y_0\,{\bf M}^T\,\y_0^T\\
          &={\bf M}\,{\bf S}(0)\,{\bf M}^{-1}\\
\label{eq_matched}
}
Hence the evolution of the matrix ${\bf S}$ in time is described
by a symplectic similarity transformation, i.e. the eigenvalues of
${\bf S}$ are conserved (which follows already from Lax' theorem).

A typical problem in accelerator physics requires to determine
a so-called {\it matched} beam: Given that the (symplectic)
transfer matrix ${\bf M}$ of some beamline (or accelerator ring)
is known, the problem is to find a stable (``matched'') distribution 
${\bf S}$ for given beam emittances. Matrices that share a system of 
eigenvectors, commute with each other. Hence, provided that ${\bf S}$
and ${\bf M}$ share a system of eigenvectors, we obtain from 
Eq.~\ref{eq_matched}:
\begeq
{\bf S}(t)={\bf S}(0)\,.
\endeq
Hence, in the matched case, the same transformations that allow
to block-diagonalize (the Hamiltonian part of) the matrix ${\bf M}$, 
also diagonalize ${\bf S}$. Provided one has an algorithm
to generate single-variate Gaussian distributions, then the described
block-diagonalization enables to generate arbitrary multivariate matched 
distributions by application of the reverse transformation~\cite{stat_paper}.

\section{Summary and Outlook}

The described Jacobi method allows to bring (skew-) Hamiltonian and
symplectic matrices to block-diagonal form by means of real symplectic
(canonical) transformations. This enables to compute 
matrix exponentials and logarithms, eigenvalues and eigenvectors of
Hamiltonian matrices with eigenvalues on the real and/or the imaginary 
axis.

The presented analysis of the structure of Hamiltonian phase spaces
reveils a morphism between four-component spinors and points in 
a classical Hamiltonian phase space for two degrees of freedom. 
It is this same structure that has been described as 
the ``square root of geometry'' and which is also used in 
quantum-electrodynamics~\footnote{
``No one fully understands spinors. Their algebra is formally 
understood but their general significance is mysterious.  
In some sense they describe the ‘square root’ of geometry
and, just as understanding the square root of -1  took  centuries,  
the  same  might  be  true of spinors''~\cite{Farmelo}.}.

It is known (though not well-known) since long that the symplectic 
group $Sp(4)$ and the algebraic structure of the Dirac algebra are 
closely related~\cite{Dirac63}, but few authors made use of this
remarkable fact. We emphasize that the construction of both the 
real Pauli as well as the real Dirac algebra was derived exclusively 
from the symmetry properties of a general classical Hamiltonian 
phase space. It follows that there is a morphism between the 
restricted Lorentz group is and the symplectic group. 
Since the presented algorithm is based solely on real Hamiltonian 
$4\times 4$-matrices, it allows for the -- from a conceptional 
point of view -- simplest possible form of the Lorentz 
transformations as shown in Ref.~\cite{lt_paper}.

\vspace{3mm}

\begin{acknowledgments}
\LaTeX\ has been used to write this article, 
Mathematica\textsuperscript{\textregistered} has been used for
parts of the symbolic calculations. 
\end{acknowledgments}

\begin{appendix}

\section{The Real Dirac Matrices}
\label{app_rdm}

In this article we used the following representation:
\begin{equation*}
\begin{array}{rclp{5mm}rcl}
\y_0&=&\bmtx{cccc}
 & 1& & \\
-1 & & & \\
 & & & 1\\
 & & -1& \\
\emtx&&
\y_1&=&\bmtx{cccc}
 &-1& & \\
-1 & & & \\
 & & & 1\\
 & & 1& \\
\emtx
\end{array}
\end{equation*}
\begin{equation*}
\begin{array}{rclp{5mm}rcl}
\y_2&=&\bmtx{cccc}
 & & & 1\\
 & & 1& \\
 & 1& & \\
 1& & & \\
\emtx&&
\y_3&=&\bmtx{cccc}
 -1& & & \\
  &1 & & \\
 & & -1& \\
 & &  &1 \\
\emtx
\end{array}
\end{equation*}
However, any representation allows for an equivalent
algorithm, firstly due to Pauli's fundamental theorem
of the Dirac matrices, but secondly since a different 
choice of matrices mostly stems from a different 
ordering of the canonical coordinates and momenta in 
$\psi$.
\end{appendix}

\bibliography{jacobi_paper.bib}{}
\bibliographystyle{unsrt}

\end{document}